\documentclass[11pt]{article}
\def\notfinal{}
\pdftrailerid{}
\usepackage[utf8]{inputenc}
\usepackage[T1]{fontenc}
\usepackage{lmodern}
\usepackage{amsmath,amssymb,dsfont,bm}\numberwithin{equation}{section}
\usepackage{microtype}
\usepackage{graphicx,tikz,pgfplots}
\graphicspath{{images/}}
\pgfplotsset{compat=newest}
\usepackage[hyperref,amsmath,thmmarks]{ntheorem}
\usepackage{aliascnt}
\usepackage[a4paper,centering,bindingoffset=0cm,marginpar=3cm,margin=3cm]{geometry}
\usepackage[pagestyles]{titlesec}
\usepackage[font=footnotesize,format=plain,labelfont=sc,textfont=sl,width=0.75\textwidth,labelsep=period]{caption}

\usepackage{subcaption}

\usepackage[backend=biber,maxnames=10,backref=true,hyperref=true,giveninits=true,safeinputenc]{biblatex}
\ifdefined\notfinal
\else
\fi

\DefineBibliographyStrings{english}{%
	backrefpage = {cited on page},
	backrefpages = {cited on pages},
}

\title{Development of mathematical models for quantitative OCT: A review}
\author{Peter Elbau$^1$\\{\footnotesize\href{mailto:peter.elbau@univie.ac.at}{peter.elbau@univie.ac.at}}
\and Leonidas Mindrinos$^2$\\{\footnotesize\href{mailto:leomid@central.ntua.gr}{leomid@central.ntua.gr}}
\and Leopold Veselka$^1$\\{\footnotesize\href{mailto:leopold.veselka@univie.ac.at}{leopold.veselka@univie.ac.at}}}
\date{}

\DeclareFieldFormat[report]{title}{``#1''}
\DeclareFieldFormat[book]{title}{``#1''}
\AtEveryBibitem{\clearfield{url}}
\AtEveryBibitem{\clearfield{note}}

\titleformat{\section}[block]{\large\sc}{\thesection.}{0.5ex}{}[]
\titleformat{\subsection}[runin]{\bf}{\thesubsection.}{0.5ex}{}[.]

\usepackage[pdftex,colorlinks=true,linkcolor=blue,citecolor=green,urlcolor=blue,bookmarks=true,bookmarksnumbered=true]{hyperref}
\hypersetup
{
    bookmarksnumbered,
    breaklinks,
    pdfborderstyle=,
    colorlinks,
    citecolor=[rgb]{0,.65,0},
    linkcolor=[rgb]{0,0,.5},
    urlcolor=[rgb]{0,0,.5},
    pdfauthor={Authors},
    pdfsubject={Subject},
    pdftitle={Title},
    pdfkeywords={Keywords}
}

\newpagestyle{headers}
{
	\headrule
	\sethead[\footnotesize\thepage][\footnotesize\sc Authors][]{}{\footnotesize\sc Development of mathematical models for quantitative OCT: A review}{\footnotesize\thepage}
	\setfoot{}{}{}
}
\newaliascnt{proposition}{lemma}

\aliascntresetthe{proposition}

\newaliascnt{corollary}{lemma}

\aliascntresetthe{corollary}

\newaliascnt{theorem}{lemma}

\aliascntresetthe{theorem}

\theorembodyfont{\normalfont}
\newaliascnt{definition}{lemma}

\aliascntresetthe{definition}

\newaliascnt{assumption}{lemma}

\aliascntresetthe{assumption}

\theoremstyle{nonumberplain}
\theoremseparator{:}
\theoremheaderfont{\normalfont\itshape}

\theoremsymbol{\ensuremath{\square}}

\newcommand{\N}{\mathds{N}}

\newcommand{\R}{\mathds{R}}
\newcommand{\C}{\mathds{C}}

\let\RE\Re
\let\Re=\undefined
\DeclareMathOperator{\Re}{\RE e}
\let\IM\Im
\let\Im=\undefined
\DeclareMathOperator{\Im}{\IM m}
\DeclareMathOperator{\supp}{supp}

\newcommand{\e}{\mathrm e}
\let\ii\i
\renewcommand{\i}{\mathrm i}

\usepackage{tabularx}

\usepackage{booktabs}

\begin{document}

\maketitle

\thispagestyle{empty}
\begin{center}
\footnotesize$^1$Faculty of Mathematics, University of Vienna, Austria

\footnotesize$^2$Department of Mathematics, National Technical University of Athens, Greece
\end{center}

\begin{abstract}
We review mathematical models describing how Optical Coherence Tomography works. Hereby, we focus on models based on Maxwell's equations and their simplifications. We highlight especially the effects of different modeling assumptions for the incident illumination, the medium, the light propagation, and the measurement setup and illustrate the qualitatively differing behavior in numerical simulations of the OCT data and compare them with real data from OCT measurements.
\end{abstract}

\section{Introduction}\label{sec_intro}

Optical Coherence Tomography (OCT) is a non-invasive imaging modality using non-ionizing radiation, typically light in the near infrared spectrum. Even though the first OCT imaging system was built only thirty years ago \cite{HuaSwaLinSchuStiCha91}, it has attracted much interest and it is considered nowadays as a key imaging technique, especially in ophthalmology, since it can acquire images very quickly (around $50\,000$ pixel points per second) with a high resolution of a few micrometers and a tissue penetration depth of approximately two centimeters. In fact, there exist more than $70\,000$ research papers related to OCT, of which around $50$ are discussing a concrete mathematical model describing how an OCT system works and how the measured data are related to the sample under investigation.

In this work, we aim to give a brief overview of the different approaches to model the various OCT setups.
Hereby, there are many modeling questions to be answered before formulating an appropriate model: How is the sample illuminated? Is there light absorption in the medium? Does the polarization of the light play an important role or can it be neglected? Do we take multiple scattering effects into account, or is it enough to focus on single scattering? What kind of detectors are being used? Just to name a few relevant decisions to be made. 

The basis of our analysis is the mathematical description presented in the review paper \cite{Fer96} by one of the inventors of OCT, Adolf Fercher, together with \cite{ElbMinSch15} where a general formulation in this spirit was presented. We will thus describe the light as an electromagnetic wave and will be using Maxwell's macroscopic equations as our model for the light propagation, which reduces in this setting to a Helmholtz equation for the electric field. We will then start by giving an overview of some of the most prominent OCT setups and the corresponding modeling assumptions in this setting in Section~\ref{se:model}, and show examplarily some of the resulting forward models for OCT in Section~\ref{se:examples}. In Section~\ref{sec_examples}, we will present the effects of the different assumptions and simplifications to the forward simulations and compare the results to real measurement data.

\section{Different Modeling Choices for OCT}\label{se:model}

There exist many different variations in the experimental setup of OCT, but its basic working principle is to illuminate the sample from a fixed direction (we will use the third basis vector $e_3$ as the illumination direction) with an incident light beam $E_I$ and to measure the backreflected light $E_S$ by interferometry, that is, to superimpose $E_S$ with a copy $E_R$ of the incident field (typically produced by reflecting $E_I$ at a mirror) and to detect the resulting intensity.

The light propagation is thereby commonly described as an electromagnetic wave, which is a pair of an electric and a magnetic field evolving according to Maxwell's macroscopic equations. Since all the media the light is interacting with are typically assumed to be non-magnetic, it is sufficient to keep track of the electric field $\hat E\colon\R\times\R^3\to\R^3$ and the electric displacement field $\hat D\colon\R\times\R^3\to\R^3$, which are functions of time and space; although it is usually more convenient to do the calculations in Fourier domain, that is, for the functions
\begin{align*}
&E\colon\R\times\R^3\to\C^3,\;E(k,x)=\int_{\R}\hat E(t,x)\e^{i kct}dt,\quad\text{and}\\
&D\colon\R\times\R^3\to\C^3,\;D(k,x)=\int_{\R}\hat D(t,x)\e^{i kct}dt,
\end{align*}
where the constant $c$ denotes the speed of light in vacuum.

\subsection{The Illumination}
The incident light is a solution of Maxwell's equations in the vacuum and can be seen as the electric field which would be produced in the absence of the sample. The equations can be in Fourier space reduced to the Helmholtz equation
\begin{equation}
\label{eq:helmholtz_vac}
\Delta E_I(k,x) + k^2 E_I(k,x) = 0, \quad (k,x)\in\R\times\R^3,
\end{equation}
for the electric field $E_I$ of the incident light with the additional condition that
\begin{equation}
\label{eq:divFree}
\nabla\cdot E_I(k,x)=0,
\end{equation}
where the differential operators $\nabla$ and $\Delta$ only act on the spatial variable $x$.

\begin{description}
\item[Swept Source OCT vs.\ Broadband Illumination:]
In swept source OCT, the incident wave is almost monochromatic, meaning that we have for every $x\in\R^3$ that
\[ \supp E_I(\cdot,x)\subset[-k_0-\varepsilon,-k_0+\varepsilon]\cup[k_0-\varepsilon,k_0+\varepsilon] \]
for a sufficiently small $\varepsilon>0$, and the measurement is repeated for multiple values $k_0$ in some interval $[k_{\text{min}},k_{\text{max}}]$, $0<k_{\text{min}}<k_{\text{max}}$. (We remark that the function $E_I$ in Fourier space fulfills the symmetry $E_I(-k,x)=\overline{E_I(k,x)}$, since each component of the electric field $\hat E_I$ is real-valued.)

Alternatively, the illumination can be done with one illumination with a broadband laser beam with
\[ \supp(\sup\nolimits_{x\in\R^3}|E_I(\cdot,x)|) = [-k_{\text{max}},-k_{\text{min}}]\cup[k_{\text{min}},k_{\text{max}}]. \]

Since we will see that we can solve the equations for the resulting electric field $E(k,\cdot)$, in particular the vector Helmholtz equation \eqref{eq:vector_helm}, for each value of $k$ separately, this difference is irrelevant for the mathematical model and we formulate it in the following therefore simply for a single broadband illumination.

\item[Gaussian Beams:]
In the classical OCT setup, the illumination is done by a laser beam (propagating along the negative $e_3$ direction) which is focused on a certain spot $r\in\R^3$, which is usually chosen inside the sample. To record the full image, measurements are done for illuminations at every horizontal position $(r_1,r_2)\in\R^2$ (practically, on some finite grid). The depth $r_3$ of the focus is thereby normally kept constant. Thus, the incident light is typically well described by a Gaussian shaped beam, see, for example, \cite{Sve10} for a treatise of the generation of light with a laser.

To obtain an expression for such a beam, we may solve the equations \eqref{eq:helmholtz_vac} and \eqref{eq:divFree} for an arbitrary value $k$ with initial data of the form
\begin{equation}\label{eq:incident_plane}
\tilde E_I(k,r_1+x_1,r_2+x_2,r_3) = f_k(x_1,x_2) \eta_k, \quad (x_1,x_2)\in\R^2,
\end{equation}
in the focus plane $\{x\in\R^3\mid x_3=r_3\}$, where $f_k\colon\R^2\to\C$ is a (Gaussian like) function such that its two-dimensional Fourier transform $\hat f_k$ is supported in the disk $D_{|k|}(0)\subset\R^2$ with center zero and radius $|k|$ and $\eta_k\in\R^2\times\{0\}$ denotes a fixed polarization vector. This has a solution which is symmetric with respect to $x_3$ regarding the focus plane of the form
\[ \tilde E_I(k,r+x) = \tilde E_{I,+}(k,r+x)+\tilde E_{I,-}(k,r+x) \]
with
\[ \tilde E_{I,\pm}(k,r+x) = \frac{1}{8\pi^2} \int_{D_{|k|}(0)} \hat f_k(\kappa)\tilde\eta_{k,\pm}(\kappa) e^{\pm i kx_3\sqrt{1-\frac{|\kappa|^2}{k^2}}} e^{i(\kappa_1 x_1+\kappa_2x_2)} d \kappa \]
and the adapted polarization vectors
\[ \tilde\eta_{k,\pm}(\kappa) = \left(\eta_{k,1},\eta_{k,2},\pm\frac{\eta_{k,1}\kappa_1+\eta_{k,2}\kappa_2}{\sqrt{k^2-|\kappa|^2}}\right). \]

However, since the incident wave should be propagating downwards, so that before a certain time, say $t=0$, we can ensure that $\hat E_I(t,r+x)=0$ for all points $x$ below a certain horizontal plane where the sample is located, say in the half space $\{x\in\R^3\mid x_3<0\}$; we know for every $x\in\R^3$ with $x_3<0$ that the function $k\mapsto E_I(k,r+x) = \int_0^\infty\hat E_I(t,r+x)e^{i kct}dt$ must be holomorphically extendable to the upper half complex plane such that $\omega\mapsto\int_\R |E_I(k+i\omega,r+x)|^2dk$ is bounded, the function $E_I$ should be only a combination of the terms $e^{i k\phi\cdot x}$ with $\phi\in\mathds S^2$ and $\phi_3<0$ (corresponding to the plane waves $(t,x)\mapsto e^{i k(\phi\cdot x-ct)}$ moving in the direction $\phi$).

Thus, we could keep as the initial wave the downwards moving part $E_I=\tilde E_{I,-}$ of our symmetric solution, where we assume $f_k=0$ for $|k|\notin[k_{\text{min}},k_{\text{max}}]$ and impose the reality assumptions $f_{-k}=\overline{f_k}$ and $\eta_{-k}=\eta_k$. If we consider the typical case that $\hat f_k$ has a very small support around the origin, where we want to assume the Gaussian behavior
\begin{equation}
\label{eq:focus_func}
\hat f_k(\kappa) \approx A_ke^{-a_k|\kappa|^2} 
\end{equation}
for some parameters $A_k\in\C$ and $a_k>0$, we can keep only the lowest order in $\kappa$ in the integrand and find the classical form
\begin{equation}\label{eq:GaussianClassic}
\begin{split}
E_I(k,r+x) &\approx \frac{A_k\eta}{8\pi^2} \int_{\R^2} e^{-a_k|\kappa|^2} e^{-i(k-\frac{|\kappa|^2}{2k})x_3} e^{i(\kappa_1 x_1+\kappa_2x_2)} d \kappa \\
&= \frac{A_kk\eta}{4\pi(2a_kk-ix_3)}e^{-ikx_3}e^{-\frac{k}{4a_kk-2ix_3}(x_1^2+x_2^2)}
\end{split}
\end{equation}
for a Gaussian beam, depicted in Figure~\ref{fig:Gaussian}.

\begin{figure}[htb]
\centering 
\includegraphics[width = 0.8\textwidth, height=0.33\textwidth]{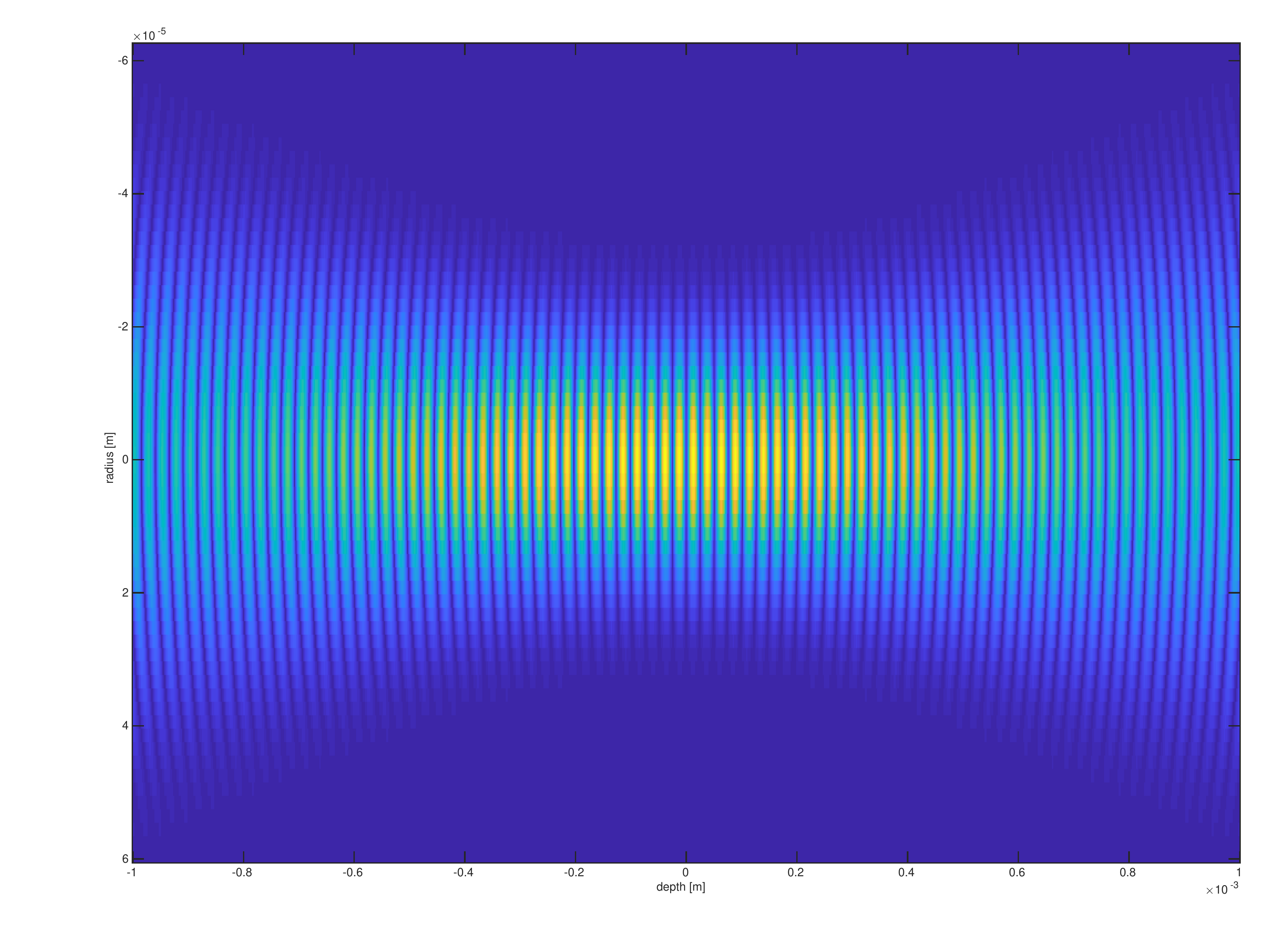}
\caption{A focused Gaussian beam presenting various forms of wavefronts.}
\label{fig:Gaussian}
\end{figure}

\item[Plane Waves:]
Approximating the expression \eqref{eq:GaussianClassic} close to the focus point, that is, for $x\approx 0$, we see that the shape of the Gaussian beam resembles the one of the plane wave
\[ E_I(k,r+x)\approx\frac{A_k\eta}{8\pi a_k}e^{-ikx_3}. \]
Thus, if the object is placed at the beam waist and its size is (for every $k\in[k_{\text{min}},k_{\text{max}}]$) of the same order of magnitude as the Rayleigh length $2a_kk$ of the light beam, then we can approximate the illumination by a plane wave, see \cite{Fer96}. Since this assumption simplifies the mathematical analysis a lot, most of the works that followed used this form of illumination, see, for example, \cite{BreReiKie16, BruCha03, BruCha05, Fer10, FerDreHitLas03, IzaCho08, KimChoParShi18, TomWan06} and some papers from the authors of this article \cite{ElbMinSch17, ElbMinSch15, ElbMinSch18a, ElbMinVes21}.

However, in particular, intensity changes due to the distance to the focus point are not reflected in this approximation which can have a negative impact for quantitative reconstructions.

Besides this rastering of the sample with focused illuminations, there also exist full-field OCT methods where the whole sample is illuminated at once. For these setups, the illumination is typically well modeled by plane waves, see \cite{MarDavBopCar09,MarRalBopCar07}, for example.

\end{description}

\subsection{The Medium}
In Maxwell's macroscopic equations, the medium is usually modeled as a non-magnetic, dielectric medium, whose optical properties are characterized by its electric permittivity $\epsilon\colon\R\times\R^3\to\C^{3\times 3}$, a function of the wavenumber $k$ and the position $x$, which relates the electric displacement field $D$ and the electric field $E$ in the medium via the linear relationship
\begin{equation}\label{eq:permittivity}
D(k,x)=\epsilon(k,x)E(k,x).
\end{equation}

In the end, we are interested in the inverse problem of reconstructing $\epsilon$ from the OCT data; however, this data seems not sufficient for a full recovery of the nine four-dimensional, complex-valued entries of $\epsilon$. It is therefore common to restrict the analysis to media with a certain structure where a reconstruction becomes possible.

There are various popular assumptions, which can in different combinations be suitable.

\begin{description}
\item[Isotropic Medium:]
In general, the inner structure of the dielectric medium can cause that the displacement field $D$ produced in the presence of an electric field $E$ is not parallel to $E$ so that the permittivity $\epsilon$ is indeed matrix-valued. To recover in this case some of the matrix entries, there exists the so-called polarization sensitive OCT method which makes illuminations with differently polarized incident waves and also detects the polarization state of the reflected light, see \cite{HeeHuaSwaFuj92}. At least for orthotropic media (that is, media whose permittivity $\epsilon$ fulfill that $\epsilon(k,x)$ is symmetric and $\epsilon_{13}=0=\epsilon_{23}$), a functional OCT scheme has been addressed in \cite{ElbMinSch18a}.

Mostly, however, it is assumed that the medium is isotropic, meaning that $\epsilon$ is a multiple of the identity matrix:
\[ \epsilon(k,x)=n^2(k,x)\mathds I_{3\times3} \]
with the refractive index $n\colon\R\times\R^3\to\{z\in\C\mid\Im z\ge0\}$.

\item[Non-Absorbing Medium:]
For biological samples in the near-infrared spectrum (that is, with central wavelengths between $800$ and $1300$ nanometers), where OCT typically operates, we have that the scattering effects dominate the absorption effects. Thus, it is often a reasonable assumption to neglect absorption completely, which means to assume that $\epsilon(k,x)\in\R^{3\times3}$ for all $(k,x)\in\R\times\R^3$.

\item[Non-Dispersive Medium:]
In addition, the bandwidth of the light source in a typical OCT system is within a narrow spectrum of a few tens of nanometers. Thus, it is commonly assumed that the permittivity is independent of the wavenumber $k$ over that small bandwidth of the source, see \cite{BruCha03, BruCha05, RajSchrAmj17}, for example. However, this way dispersion effects are excluded from the model.

\item[Slowly Varying in Horizontal Direction:]
Since the laser illuminates only a small region inside the sample, we can assume that the properties of the medium within the illuminated region are almost independent of the lateral coordinates $x_1$ and $x_2$, so that the permittivity can be locally considered to depend spatially only on the depth $x_3$.

\item[Weakly Scattering Medium:]
An assumption with mainly theoretical motivation, is to pretend that the medium is only weakly scattering, meaning that $\epsilon-\mathds I_{3\times3}$ is so small that we can apply the Born approximation to linearize the forward problem and get an explicit formula for it, see \cite{ElbMinSch15, Fer96, SchmKnuBon93, TriCar19}. Allowing for strongly scattering media in which also multiple scattering events needs to be considered, results in contrast in more complicated models \cite{AduHilTim07, KarLauLeuBouLas05, NguFabVan13, YadSchmBon95}.

\item[Layered Medium:]
One of the main applications of OCT is in ophthalmology where the retina of the human eye is imaged. Another important application is the imaging of human skin. Both of them are biological tissues presenting a layered structure, where each layer is described by a distinct permittivity (usually with additional simplifying assumption such as being constant or only weakly scattering) so that we have disjoint bounded sets $\Omega_j\subset\R^3$, $j\in\{1,\ldots,J\}$, with a smooth boundary and local permittivities $\epsilon_j\colon\R\times\R^3\to\C^{3\times3}$, $j\in\{1,\ldots,J\}$, for which the permittivity has the form
\[ \epsilon(k,x) =  \bm \chi_{\R^3 \setminus \Omega}(x) \mathds I_{3\times3} + \sum_{j=1}^J \bm \chi_{\Omega_j}(x) \epsilon_j(k,x), \]
where $\Omega=\bigcup_{j=1}^J\Omega_j$ is the region of the sample and $J\in\N$ denotes the number of layers, see, for example, \cite{ElbMinSch15, FerHitStiZaw02, FerHitSti01, Kal17}.

Often one makes the additional assumption that, at least locally in the illuminated region, the domains $\Omega_j$ can be approximated by the sets
\[ \{x\in\R^3\mid z_{j+1}<x\cdot\nu_{\Omega,j+1},\,x\cdot\nu_{\Omega,j}\le z_j\} \]
which are bounded by planes with upwards pointing normal unit vectors $\nu_{\Omega,j}\in\mathds S^2$ and parameters $z_j\in\R$ (all depending on the position of the illuminated region).

\item[Point-Like Scatterers:]
Similarly to the layered media, one may assume the inhomogeneities inside the object as randomly distributed point-like scatterers with constant optical properties. Then, the aim is to recover the geometry of the area where the scatterers are located \cite{Mun16, RalMarCarBop06, SchmXiaYun99, YunLeeSchm99}.

\end{description}

\subsection{The Scattering}
In general, Maxwell's macroscopic equations describe perfectly the wave propagation into the imaging system and the scattering theory is the main tool to derive a formula for the backscattered light from the object, see, for example, \cite{ColKre13, ManWol95}. Since most OCT systems are using focused illuminations and are thus essentially performing multiple one-dimensional measurements, the electromagnetic theory can be considerably simplified, such as in \cite{AndThrYurTycJorFro04, ElbMinSch15, Fer10, FerDreHitLas03, SchmKnu97, SilCor13, ThrYurAnd00}, to name but a few examples.

In the general case, Maxwell's macroscopic equations can be (implementing the medium via the relation \eqref{eq:permittivity}) combined into the vector Helmholtz equation
\begin{equation}
\label{eq:vector_helm}
\nabla\times(\nabla\times E)(k,x) = k^2\epsilon(k,x) E(k,x),\quad k\in\R,\;x\in\R^3.
\end{equation}
for the electric field $E\colon\R\times\R^3\to\C^3$, which is induced by the incident field $E_I$, that is, we request an initial condition of the form $\hat E(t,x)=\hat E_I(t,x)$ for all times $t<t_0$ with the time $t_0$ be chosen such that $\hat E_I$ is not interacting with the sample before $t_0$: $\hat E_I(t,x)=0$ for all $t<t_0$ and $x\in\bigcup_{k\in\R}\supp(\epsilon(k,\cdot))$. Choosing for simplicity $t_0=0$, this can be again formulated as a boundedness condition on the analytic continuation of the inverse Fourier transform $E_S$ of the scattered fields $\hat E_S=\hat E-\hat E_I$ on the upper half complex plane (as indicated in the derivation of \eqref{eq:GaussianClassic}), or more commonly expressed by the Silver--Müller radiation condition
\[ \lim_{|x|\to\infty}((\nabla\times E_S(k,x))\times x-ik|x|E_S(k,x)) = 0, \]
see \cite{ColKre13}, for example.

The differential equation \eqref{eq:vector_helm} together with this initial condition can then be rewritten as the Lippmann-Schwinger integral equation
\begin{equation}
\label{eq:lippmann}
E(k,x)-E_I(k,x) = \mathcal U[E](k,x)
\end{equation}
with the integral operator $\mathcal U$ given by
\[ \mathcal U[E](k,x)=\left( k^2\mathds I + \nabla \nabla\cdot\right)\int_{\R^3} \Phi(k,x-y) (\epsilon(k,y)-\mathds I_{3\times3}) E(k,y) d y, \]
where $\mathds I$ is the identity operator and  
\[
\Phi\colon\R\times(\R^3\setminus\{0\})\to\C,\quad \Phi(k,x) = \frac{e^{ik|x|}}{4\pi|x|},
\]
denotes the three-dimensional outgoing Green's function of the Helmholtz operator (that is, the solution of the distributional equation $\Delta\Phi+k^2\Phi=-\delta$ fulfilling the boundedness condition for $k$ in the upper half complex plane). We note that the integration area in the operator $\mathcal U$ is restricted to the support of the functions $\epsilon(k,\cdot)-\mathds I_{3\times3}$, which represents the sample and should thus be bounded.

\begin{description}
\item[Born Approximation:]
The integral \eqref{eq:lippmann} is only an implicit equation for the scattered field. If we make the assumption of a weakly scattering medium, meaning that we consider the deviation $\epsilon(k,y)-\mathds I_{3\times3}$ of the permittivity from the permittivity of free space to be small, we may neglect in a first order approximation the second order term $(\epsilon(k,y)-\mathds I_{3\times3})(E(k,y)-E_I(k,y))$ in the integrand and approximate $E_S$ by its first order Born approximation, given by the explicit expression
\[ E_S^{(1)}(k,x) = \mathcal U[E_I](k,x). \]

The product of $(\epsilon(k,y)-\mathds I_{3\times3})E_I$ at a point $y$ can be interpreted as the response of the medium there to the (unperturbed) incident illumination $E_I$, which acts as a point source that propagates according to the Green's function to an examination point $x$. The integral over the object volume then provides the total contribution of the sample.

However, the approximation does not reflect that these additional contributions also interact with the medium. This could be iteratively taken into account by using higher order Born approximations defined by
\[ E^{(\ell+1)}_S(k,x) = \mathcal U[E_I+E^{(\ell)}_S](k,x),\quad\ell\in\N. \]

But although the usage of higher-order Born approximation in case of a strongly deviating refractive index is possible, it is due to the evaluation of many integrals not very efficient. Therefore, other possibilities such as the implementation of the radiative transfer equation and Monte Carlo simulations are often preferred, see \cite{GelDolSerTur03, KarLauLeuBou05, KarLeuLauBou05b,  NguFabVan13, TeaBreSou95, YadSchmBon95}.

\item[Far-Field Approximation:]
Maybe less an approximation, but rather attributed to the comparably large distance of the detector to the sample, it is sometimes considered that the scattered field there is only known in highest order so that barely the first term $E_S^{(\infty)}(k,\tfrac x{|x|})$ in the expansion
\[ E_S(k,x) = \frac{e^{ik|x|}}{4\pi|x|}\left(E_S^\infty(k,\tfrac x{|x|})+\mathcal O(\tfrac1{|x|})\right) \]
for $|x|\to\infty$ significantly enters into the measurements.

In particular in combination with the Born approximation, this nevertheless simplifies the expression for the scattered field as we end up with 
\[ E^{(1,\infty)}(k,\vartheta) = -k^2\int_{\R^3}  e^{-ik \vartheta \cdot y} \vartheta \times \Big( \vartheta \times \big(\epsilon(k,y)-\mathds I_{3\times3}\big)E_I(k,y)\Big) dy, \]
which makes it a very common model, see, for example, \cite{ElbMinSch15, Fer96, Fer10, FerDreHitLas03, FerHit02, SanAra15}, especially, combined with plane wave illumination. When it comes to Gaussian beam illumination, the difference between far- and near-field representation becomes more noticeable. The latter counts also for system effects that are neglected when the far-field approximation is applied, see \cite{BreReiKie16, ElbMinVes21, YiLi09}.

\item[Linear Polarization:]
If we assume an isotropic medium described by a refractive index $n$, then we can write the vector Helmholtz equation \eqref{eq:vector_helm} (using the vector identity $\nabla\times(\nabla\times E)=\nabla(\nabla\cdot E)-\Delta E$ together with the direct consequence $\nabla\cdot(n^2 E)=0$ of \eqref{eq:vector_helm}) in the form
\[ \Delta E(k,x) + k^2 n^2(k,x) E(k,x) +\nabla\left(E\cdot \nabla\left(\ln n^2(k,x)\right)\right) = 0, \quad k\in\R,\;x\in\R^3, \]
from which we see that if $E\cdot\nabla n$ vanishes identitcally, then the vector Helmholtz equation is reduced to the scalar Helmholtz equation for every component of $E$. This is, for example, the case if $E_I(k,x)=u_I(k,x)\eta$ with $\eta\in\R^2\times\{0\}$ and the refractive index $n$ is assumed to depend only on $x_3$ (which may be approximated by assuming a medium which is slowly varying in horizontal direction) so that $\nabla n$ is parallel to $e_3$. Then, the electric field is of the form $E(k,x)=u(k,x)\eta$ with $u$ being a solution of
\[ \Delta u(k,x)+k^2n^2(k,x)u(k,x)=0, \]
thus reducing the vector-valued problem to a scalar one, see, for example, \cite{AndThrYurTycJorFro04,Fer96,Fer10,FerDreHitLas03,MarRalBopCar07}.

\item[Fresnel Formulas and the Huygens--Fresnel Principle:]
We consider for simplicity a layered medium whose layers are bounded by parallel planes perpendicular to an upwards pointing normal vector $\nu_\Omega\in\mathds S^2$ where the local permittivities are assumed to be isotropic, non-dispersive and spatially constant: $\epsilon_j(k,x)=n_j^2\mathds I_{3\times3}$. Then, the electric field $E$ in each layer $\Omega_j$ will be given as a superposition of plane waves of the form
\[ E_{j,\phi,\eta}(k,x)=A_{j,\phi}e^{ikn_j\phi\cdot x}\eta_{j,\phi} \]
with different propagation directions $\phi\in\mathds S^2$ for appropriately chosen amplitudes $A_{j,\phi}\in\C$ and polarization vectors $\eta_{j,\phi}\in\mathds S^2$ with $\phi\cdot\eta_{j,\phi}=0$.

The amplitudes and polarizations have then to be chosen so that at every point $x_{\Omega,j}\in\partial\Omega_j\cap\partial\Omega_{j+1}$ on a discontinuity of the refractive index, the transmission conditions
\begin{align*}
&\lim_{m\to\infty}E(k,\bar y_m) = \lim_{m\to\infty}E(k,\underline{y}_m)\quad\text{and} \\
&\lim_{m\to\infty}n_jE(k,\bar y_m)\cdot\nu_\Omega = \lim_{m\to\infty}n_{j+1}E(k,\underline{y}_m)\cdot\nu_\Omega
\end{align*}
are fulfilled for all sequences $(\bar y_m)_{m=1}^\infty\subset E_j$ and $(\underline{y}_m)_{m=1}^\infty\subset E_{j+1}$ with $\lim_{m\to\infty}\bar y_m=x_{\Omega,j}$ and $\lim_{m\to\infty}\underline{y}_m=x_{\Omega,j}$.

For only two (infinite) layers and an incident plane wave, the solution can be shown to be of the form
\[ E(k,x) = \bm\chi_{\Omega_1}(x)\Big(Ae^{ikn_1\phi\cdot x}\eta+\tilde Ae^{-ikn_1\phi\cdot x}\tilde\eta\Big)+\bm\chi_{\Omega_2}(x)Be^{ikn_2\tilde\phi\cdot x}\bar\eta, \]
where the amplitudes $\tilde A,B\in\C$, the polarization vectors $\tilde\eta,\bar\eta\in\mathds S^2$, and the propagation direction $\tilde\phi\in\mathds S^2$ of the reflected and transmitted waves can be explicitly calculated via the Fresnel equations from the amplitude $A\in\C$, the polarization $\eta\in\mathds S^2$ and the direction $\phi\in\mathds S^2$ of the incident plane wave, see, for example, \cite{Jac98}.

Now, similarly to the higher Born approximation, we can iteratively calculate the solution of the general scattering problem by---starting with the incident wave, decomposed as a sum of plane waves---successively reflecting and scattering the different plane waves at the discontinuities of the layers. Since each reflection step goes along with a loss in amplitude, one can in practice only keep those plane wave components which were produced from a low number of reflections. For a more detailed explanation of the construction, we refer to \cite{BruCha05, ElbMinVes21, Schm99}.

We want to mention at this point that even if we take in this construction only contributions from waves into account which have been reflected at most once, the so obtained single scattering solution would be different from the corresponding Born solution, since it would still depend non-linearly on the refractive index (as it appears in the exponent of every plane wave component).

We refer to \cite{BruCha03, BruCha05, HabBlaSchm98, SeeMul14} for works using similar modeling. The layered structure can be also combined with different differential equations, like the radiative transfer equation \cite{GelDolSerTur03, TurSerDolSha05}.

Instead of calculating the solution inside the layers completely, there also exists the approach of applying the extended Huygens-Fresnel principle, see \cite{BorWol99,LutYur71,YurHan87}, to propagate the values of the electric field from one layer boundary directly to the next layer boundary and in that way also back to the detector, see \cite{AndThrYurTycJorFro04, SchmKnu97,ThrYurAnd00}. 

\end{description}

\subsection{The Measurements}

Since the time resolution is usually not sufficient to determine from measurements of the rapidly oscillating electromagnetic waves $\hat E_S$ the Fourier transform $E_S$, the way to get information about the phase is to split off a part of the incident wave via a beam splitter and reflect it at a mirror, which produces a well-known electromagnetic field $E_R$ (which can be calculated from the Fresnel equations), and recombine it with the scattered field $E_S$ to measure the superposition $E_S+E_R$. There are slight variations on how this is exactly done.

\begin{description}
\item[Time-Domain vs.\ Spectral-Domain OCT:]
In time-domain OCT, the mirror to produce the field $\hat E_R$ is movable and for every shifted position $\mu\in[-\mu_0,\mu_0]$ of the mirror, we get for an induced scattered field $\hat E_S$ at a detector point $x\in\R^3$ the averaged intensity measurement
\begin{align*}
I(\mu,x)&=\int_\R|\hat E_S(t,x)+\hat E_R(t+\tfrac{2\mu}c,x)|^2dt \\
&=\int_\R|\hat E_R(t+\tfrac{2\mu}c,x)|^2dt+2\int_\R\hat E_S(t,x)\cdot\hat E_R(t+\tfrac{2\mu}c,x)dt+\int_\R|\hat E_S(t,x)|^2dt.
\end{align*}
The dominant first term is simply the intensity of the incident light beam, which is known, and the last term can be obtained by an intensity measurement of the scattered field without superimposing it with the reference field $\hat E_R$ (but often it is simply ignored as it is of second order in the field $\hat E_S$, which is usually much weaker than $\hat E_R$). Therefore, we can assume to know by this the second, cross-correlation term.

Using the Parseval--Plancherel identity, we can write it in the form
\[ \int_\R\hat E_S(t,x)\cdot\hat E_R(t+\tfrac{2\mu}c,x)dt = \frac c{2\pi}\int_\R E_S(k,x)\cdot \overline{E_R(k,x)}e^{2i\mu k}dk, \]
so that we can calculate from this by doing a Fourier transform with respect to $\mu$ (provided the interval of $\mu$ is sufficiently large---in general, a convolution operator will additionally appear, see \cite{FerDreHitLas03}) the product
\begin{equation}
\label{eq:cc_term}
 M(k,x) = E_S(k,x)\cdot \overline{E_R(k,x)}\quad\text{for all}\quad k\in[k_{\text{min}},k_{\text{max}}]. 
\end{equation}

The expression $\Re(M)$ can alternatively be directly measured by sending the sum $\hat E_S+\hat E_R$ through a spectrometer giving the values $|E_S(k,x)+E_R(k,x)|^2$ of which the values $|E_S(k,x)|^2$ and $|E_R(k,x)|^2$ can be again obtained separately leaving only $\Re(M)$. This is the way spectral-domain OCT is working, which has the experimental advantage, that there is no more necessity of moving the mirror which speeds up the measurements.

We want to remark that if not a broadband signal is used, but a swept source method, then for each illumination with an almost monochromatic incident wave at wavenumber $k$, only one value $\Re(M(k,x))$ is recovered with this spectral method, and the full signal is obtained by repeating the procedure for every value $k\in[k_{\text{min}},k_{\text{max}}]$.

\item[Polarization Sensitive OCT:]
For a polarization sensitive setup, every measurement is repeated with an incident wave where just the polarization vectors in every Fourier component are changed. Additionally, the superposition $E_S+E_R$ is splitted and sent to different polarization filters so that we can measure the single components $\Re(E_{S,j}(k,x)\overline{E_{R,j}(k,x)})$, $j\in\{1,2,3\}$, of the cross-correlations.

\item[Focused Illumination vs.\ Full-field OCT:]
In full-field OCT, the whole sample is, as we have seen, illuminated once and we measure the resulting scattered field (in one of the previously described ways) at all points $x$ on a detector plane $\mathcal D=\{x\in\R^3\mid x_3=d\}$ above the sample.

In the classical setup, we generate for each point $(r_1,r_2,d)\in\mathcal D$ an incident wave $(k,x)\mapsto E_I(k,r+x)$ with focus in $r=(r_1,r_2,r_3)\in\R^3$ for some depth focus $r_3$ (inside the sample) and detect the hereby induced scattered field $E_S$ at the detector position $(r_1,r_2,d)$ only. Hereby, we produce at every point on the detector a measurement which contains information of the sample in a small neighborhood of the vertical line $\{(r_1,r_2,x_3)\mid x_3\in\R\}$, which is typically called an ``A-scan''. The reconstruction of the medium is thus essentially reduced to a (one-dimensional) reconstruction of the sample on each of these vertical lines from its corresponding A-scan.
\end{description}

\section{Examples of Forward Models}\label{se:examples}
To give a brief overview, what assumptions are typically made and how they are usually combined, we tried to collect and classify some of the theoretical works on OCT in Table~\ref{tb:classification}.

\begin{table}[htb]
\begin{itemize}
\item[] \textbf{Illumination}\\
p.w. = plane wave, G.b. = Gaussian beam
\item[] \textbf{Medium} \\
w.s. = weakly scattering, (n.-)abs. = (non-)absorbing, (n.-)disp. = (non-)dispersive, lay. = layered medium, p.sca. = point scatterers
\item[] \textbf{Scattering} \\ 
B.A. = Born approximation, F.A. = far-field approximation, Huy. = Huygens principle l.p. = linear polarization, FF. = Fresnel formulas        
\end{itemize}

\noindent\begin{tabularx}{\linewidth}{*{6}{X}}
\toprule
Articles &  Illumination & Medium & Scattering & Measurement \\ 
\midrule
\cite{ElbMinSch15, Fer96, Fer10, FerDreHitLas03, FerHit02, IzaCho08, Kal17, MarRalBopCar07} & p.w. & w.s., disp. & B.A. \&/or F.A.,\, l.p. & FD \&/or TD\\ \hline
\cite{BruCha03, BruCha05, ElbMinVes20, RajSchrAmj17, TomWan06} &  p.w. &  n.-abs. \&/or abs., n.-disp., lay. & l.p., FF. & FD  \\\hline 
\cite{RalMarCarBop06, VesKraMinDreElb21} & G.b. & p.sca., lay. & F.A., FF. &  FD  \\\hline
\cite{AndThrYurTycJorFro04, SchmKnu97, ThrYurAnd00} & G.b. & w.s., lay. & Huy. & TD \\
\bottomrule
\end{tabularx}

\caption{Classification of some articles on OCT depending on the considered setup.}\label{tb:classification}
\end{table}

Although we cannot discuss all of them here, we want to at least show at two particular cases how these assumptions are combined to a complete forward model.

\subsection{Single Scattering Plane Wave Model}
One of the classical models, used for example in \cite{Fer96}, assumes (locally) a plane wave illumination and considers a medium which is isotropic, non-absorbing, and non-dispersive. Then, we have seen that we can describe the medium by a refractive index $n\colon\R^3\to\C$, being a function of $x$.

Assuming further that the medium is weakly scattering and applying the Born and far-field approximations, we obtain for the far-field pattern $E_S^{(1,\infty)}$ of the scattered field $E_S$ induced by an incident plane wave $E_I(k,x)=Ae^{-ikx_3}\eta$ with $A\in\C$ and $\eta\in\R^2\times\{0\}$ in the direction $\vartheta\in\mathds S^2$ with $\vartheta_3>0$ the expression
\[ E^{(1,\infty)}(k,\vartheta) = Ak^2(\vartheta\times(\vartheta\times\eta))\int_{\R^3}\big(n^2(y)-1\big)e^{-ik(\vartheta\cdot y+y_3)} dy. \]
As expected, the measurements thus are proportional to the Fourier transform $\mathcal F[n^2-1](k(\vartheta+e_3))$ of the reflectivity $n^2-1$.  We stress here once more that OCT is a band-pass technique and we actually detect only high frequency Fourier components because only backscattered light is measured within a small frequency range. 

\begin{figure}
        \centering
        \begin{subfigure}[b]{0.475\textwidth}   
            \centering 
            \includegraphics[scale=0.75]{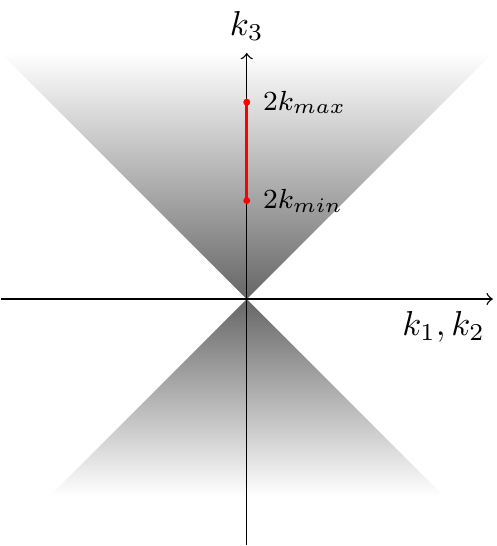}
            \caption[]%
            {{\small Standard OCT.}}    
        \end{subfigure}
        \quad
        \begin{subfigure}[b]{0.475\textwidth}   
            \centering 
            \includegraphics[scale=0.75]{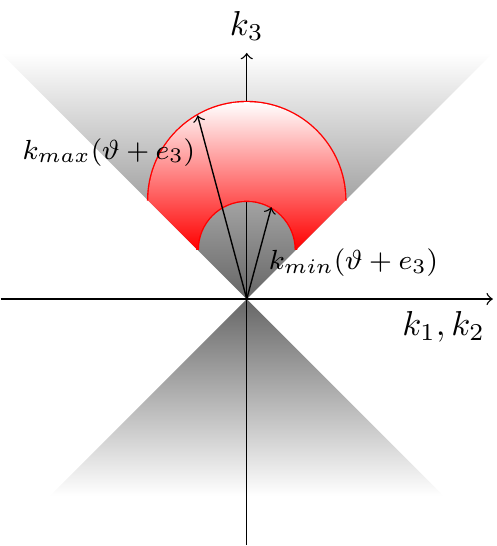}
            \caption[]%
            {{\small Full-field OCT.}}    
            \label{fig:focused}
        \end{subfigure}
        \caption[ ]{Limited Fourier data because of backscattering at few wavelengths.}
        \label{fig_fourier}
    \end{figure}

In Figure~\ref{fig_fourier} we see the few accessible Fourier data (in red) for two different OCT systems compared to the available Fourier data (gray upper cone): In standard OCT with focused illumination, we will only detect the directly back-scattered values, that is, $\vartheta=e_3$ (at each horizontal position), whereas in full-field OCT we would (ideally) obtain the contributions from all $\vartheta\in\mathds S^2$ with $\vartheta_3>0$.

This result is the main reason for the few mathematical-oriented papers, since for many years the inverse problem in OCT, to recover the reflectivity $n^2-1$ from the OCT data, was just accomplished by the application of the Fourier transform. However, the need to improve image quality and enhance quantitative analysis forced different mathematical groups to relax some of the assumptions and consider different cases to handle the more general scheme.

\subsection{A Gaussian Illumination Model}
In \cite{VesKraMinDreElb21}, we used a Gaussian incident wave $E_I$ for analyzing a layered medium with constant local refractive indices and whose layers $\{x\in\R^3\mid z_{j+1}<x\cdot\nu_\Omega\le z_j\}$ are bounded by parallel planes with normal vector $\nu_\Omega$.

Using \eqref{eq:incident_plane} for the data in the focal plane, this leads (we only consider the A-scan at the position $(r_1,r_2)=(0,0)$ to simplify the notation) to 
\begin{multline}
\label{eq:scatfield}
E_S(k,x) = \frac{1}{8\pi^2} \sum_{j=1}^{J+1} \int_{\R^2} \beta_j(k,\kappa) \left(\prod_{\ell=1}^{j-1} (1-\beta_\ell^2(k,\kappa)) e^{i k 2 n_\ell(k)(z_\ell-z_{\ell+1}) \cos \theta_{t,\ell}(k,\kappa)}\right) \hat f_k(\kappa) \\
e^{i \sqrt{k^2-|\kappa|^2}r_3} e^{i \left( K-\psi(K)\right)\cdot x_\Omega } e^{i\psi(K)\cdot x} d \kappa,
\end{multline}
where the following notation was introduced: 
\begin{itemize}
\item $\beta_j\colon\R\times\R^2\to\R^3$ denotes the dispersive reflection coefficient, accounting for the differently polarized parts of the electric field;  
\item for a $\kappa\in \R^2$ with $K = (\kappa,-\sqrt{k^2-|\kappa|^2})\in\R^3,$ we define the scattered vector
\[ \psi\colon \R^3\to\R^3, \quad K \mapsto K - 2\left( K\cdot\nu_\Omega\right) \nu_\Omega; \]
\item $\theta_{t,j}$ denotes the angle of transmission from the layer $j$ to the layer $(j+1)$;
\item $x_\Omega = (0,0,z_1)$ denotes the point on the top surface of the medium.
\end{itemize}

A part which often has been neglected in articles on OCT, is the transport of the reflected light to the detector. In practice, a common setup is that the back-reflected light is coupled into a single-mode fiber and later transferred to the detector. Since both fibers connecting the sample and the reference arm with the detector are assumed to have equal length, the additional phase factor, which commonly represents the way of light through a fiber, cancels out when being combined in \eqref{eq:cc_term}.

While the way through the fiber only takes a minor role in an OCT measurement, the coupling into the fiber has a more prominent position in here \cite{RalMarCarBop06,VesKraMinDreElb21}. This coupling is done by using a scan lens which discards all (plane wave) parts in equation \eqref{eq:scatfield} whose propagation directions strongly deviate from the illumination direction $e_3$ \cite{KnuBoe00}. The maximal deviation is described by the (maximal) angle of acceptance, which we denote by $\theta.$ The set of incident directions yielding an accepted wave vector is then given by 
\begin{equation}
\label{eq:accepwavevec}
\mathcal B_k = \left\{\kappa\in\R^2\ \Big|\ \psi(K(\kappa))\cdot e_3  \geq k\cos\theta  \right\}\subset \R^2. 
\end{equation}
Thus, we change the area of integration in \eqref{eq:scatfield} to $\mathcal B_k\subset\R^2$, which gives us (after multiplying with the reflected field) our forward model.

To simplify the expression, the integral may be replaced by its far-field approximation: For a direction $\vartheta\in \mathds S^2_+$ and (large) $\rho\in \R,$ the far-field approximation of the reflected field by a single interface, that is $j=1$ in \eqref{eq:scatfield}, is given by \cite{VesKraMinDreElb21}
\begin{equation}
\label{eq:farfield}
E_S(k,\rho\vartheta) \simeq  \frac{ i k C(\kappa_0)}{\rho} \beta(k,\kappa_0)\hat f_k(\kappa_0)  e^{i\sqrt{k^2-|\kappa_0|^2}r_3} e^{i (K-\psi(K))\cdot x_\Omega} e^{i k\rho},
\end{equation}  
where $\kappa_0$ is such that $\psi(K(\kappa_0)) = \vartheta$ and $C$ is constant depending on $\kappa_0.$

It has been shown in \cite{VesKraMinDreElb21} that the far-field approximation is meaningful only if the medium is located in the focus of the laser beam. However, there are still cases where this approximation fails as we demonstrate in Section~\ref{sec_example_far} with a numerical example.

Although we have for this model no closed form for the reconstruction of the positions $z_j$ of the layer boundaries and the refractive indices $n_j$ of the layers, the parameters can numerically be calculated to optimally fit the data which gives for suitables samples satisfying the modeling assumptions good reconstructions.

\section{Numerical Experiments}\label{sec_examples}

In this section we present numerical experiments using both simulated and real OCT data. The goal is to show that some mathematical models even if they are commonly used might not be appropriate for some specific OCT setups and some objects.

\subsection{Near- vs.\ Far-Field Representation}\label{sec_example_far}

We address a special case where near- and far-field representations of the sample field coincide at a specific point but deviate in general. 

\begin{figure}[hbt!]
\centering 
\includegraphics[width=1\textwidth]{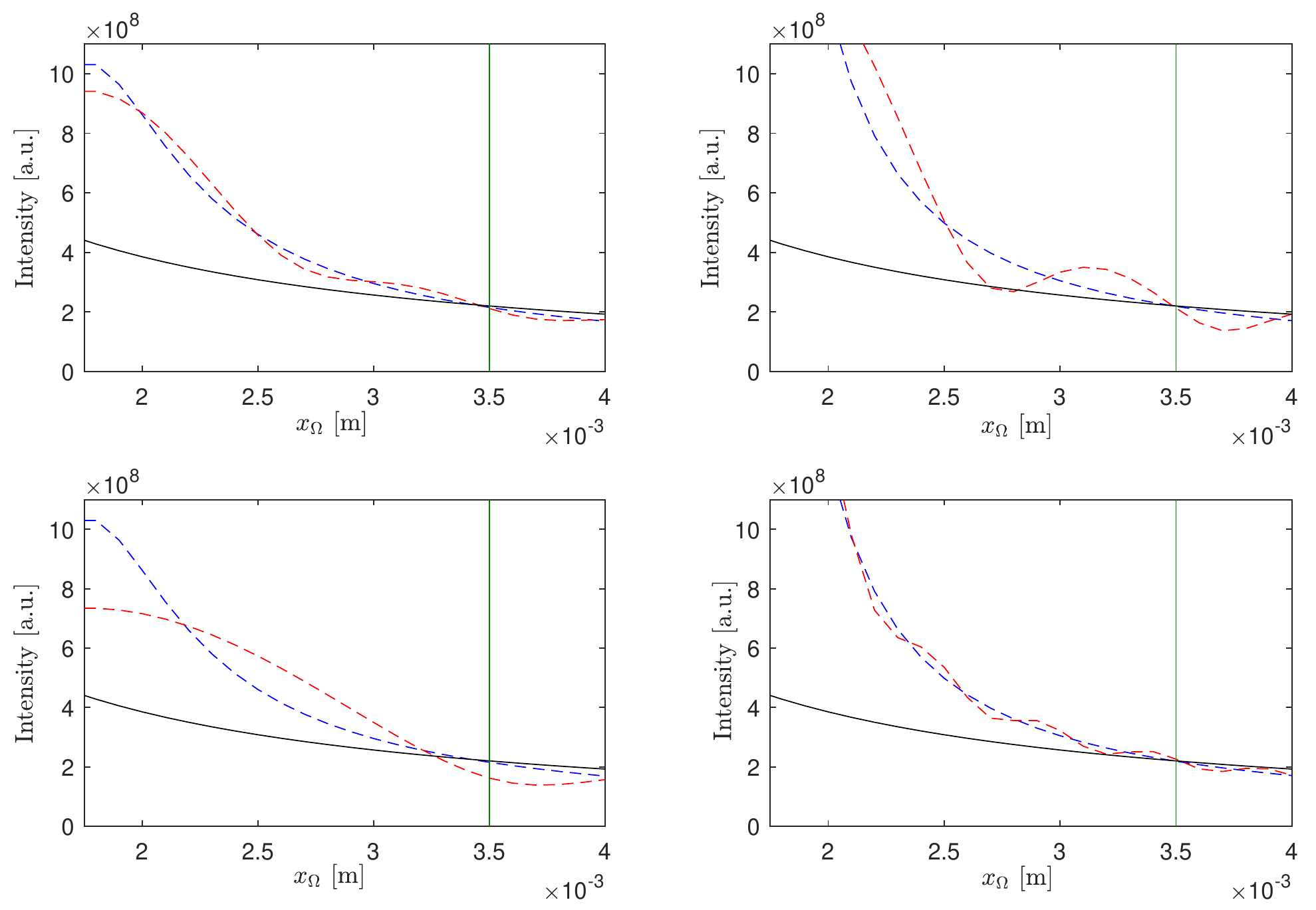}
\caption{The far-field approximation (black) in \eqref{eq:farfield}, the Gaussian near field (red) in \eqref{eq:near_field} and the Gaussian near field in the limit case $\theta =\frac\pi2$ for different sample positions $x_\Omega = (0,0,z_1).$ The green line denotes the focus position $r_3.$ The near field in red is presented for different values of $a_k$ and $\theta.$} 
\label{fig:farfield}
\end{figure}

Let the imaging object be a totally reflecting mirror that is not tilted, meaning $\beta\approx 1$ and $\nu_\Omega=e_3$.
We assume further that it is placed below the examination point $x=0.$ Then, after linearizing the root in \eqref{eq:scatfield}, we obtain the following form for the reflected field 
\[
E_S(k,0) \approx \frac{e^{i k (r_3 - 2z_1)} A_k}{8\pi^2} \int_{\mathcal B_k} e^{-|\kappa|^2 \left(a_k + \frac{i}{2k} (r_3 - 2z_1)\right)} d\kappa,
\]  
where we used the definition of $\hat f_k$ in equation \eqref{eq:focus_func}.
It is straightforward to show that $\mathcal B_k,$ in this simple case, is given by the ball with radius $k \sin(\theta).$ Thus, we parametrize the integral to get 
\begin{equation}
\label{eq:near_field}
E_S(k,0) \approx e^{i k (r_3 - 2z_1)} \frac{A_k}{8\pi\left(a_k + \frac{i}{2k} (r_3 - 2z_1)\right)} \left(1 - e^{-k^2 \sin^2(\theta)a_k} e^{-i k \left(\frac{r_3}{2} - z_1\right)}\right).
\end{equation}
For $z_1 = r_3,$ we observe that the absolute value of \eqref{eq:farfield} is only a good approximation of $|E_S(k,0)|$ if $\theta$ increases to the point that almost all back-reflected waves are collected. The difference between a small (left) and a large (right) value of $\theta$ is shown in the bottom row of Figure~\ref{fig:farfield}. 

\subsection{Plane Wave vs.\ Gaussian Shaped Illumination}\label{sec:example_gaussian}

We present an example with real data generated from a swept-source OCT system. We show that the Gaussian beam seems advantageous. In addition, we observe that even multiple reflections (included in the model) do not improve the simulations. We compare the real data with the single-reflection model for the plane and the Gaussian shaped illumination. Simulations with second-order reflections are also presented.

The OCT experiment was performed with a source centered around a wavelength $\lambda_0 \approx 1300$nm with a bandwidth from approximatively $1282.86$nm to $1313.76$nm. The medium was a three-layered sample consisting of glass, water and glass.  The sample was raster-scanned $1024\times 1024$ times in lateral directions. For each raster scan we record a spectrum consisting of $1498$ data points, equally spaced with respect to wavenumber. Due to the small bandwidth of the laser, dispersion compensation is not needed.

According to \cite{VesKraMinDreElb21}, in order to adjust all necessary parameters for the Gaussian incident field the object must be shifted multiple times along its axial direction. For each shift, the measurement process is repeated.

The optical properties of the sample are known. The used coverglass and the water have at the wavelength $1300$nm the refractive index $1.5088$ and $1.3225$, respectively. The thickness of each layer was calculated manually using the optical distance in the image space.  

\begin{figure}[hbt!]
\centering 
\includegraphics[width=1\textwidth,height=0.75\textwidth]{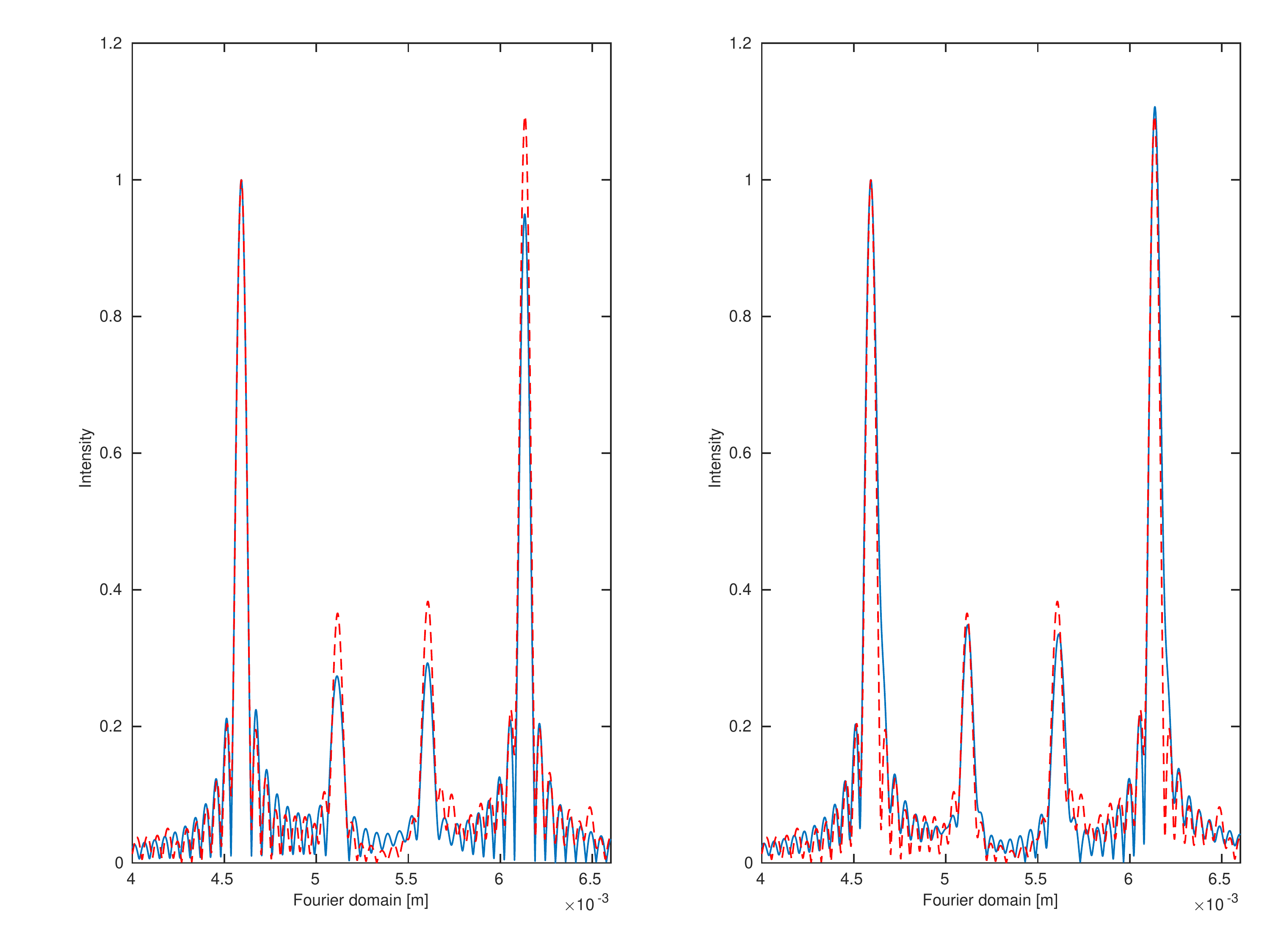}
\caption{Comparison between real data (red) and simulations (blue). On the left, the sample field was generated by a plane wave model with first order reflections. On the right, a Gaussian beam model was applied.} 
\label{fig:real_vs_plane_gaussian}
\end{figure}

In Figure~\ref{fig:real_vs_plane_gaussian} we present a comparison between the Fourier transform (with respect to wavenumber) of a single raster scan and the simulations for a plane wave and a Gaussian beam illumination.
 
The plane wave model catches better the parts away from the center of every sinc function in Figure~\ref{fig:real_vs_plane_gaussian}, which is also represented by a smaller root-mean-square error (RMSE) in Table~\ref{tb:rmse}. On the other side, the point of matching the height of the prominent peaks, the maxima of the sinc functions, is reasonably reflected only by the Gaussian model. This point however is crucial, especially when we go to the inverse problem. The quality of the reconstruction of the refractive indices is heavily dependent on the height difference between real data and forward simulations \cite{BruCha05,ElbMinVes21}. 

\begin{table}[htb]
\noindent\begin{tabularx}{\linewidth}{*{3}{X}}
\toprule
Num. of reflections & Plane Wave & Gaussian beam  \\ 
 \midrule
Single & 0.0363 & 0.0391\\
Multiple  & 0.0364 & 0.0391 \\
\bottomrule
\end{tabularx}
\caption{The RMSE between real and simulated data for the plane wave and Gaussian beam models. The error is computed for single (top row) and multiple reflections (bottom row).}
\label{tb:rmse}
\end{table}    

\begin{figure}[hbt!]
\centering 
\includegraphics[width=1\textwidth]{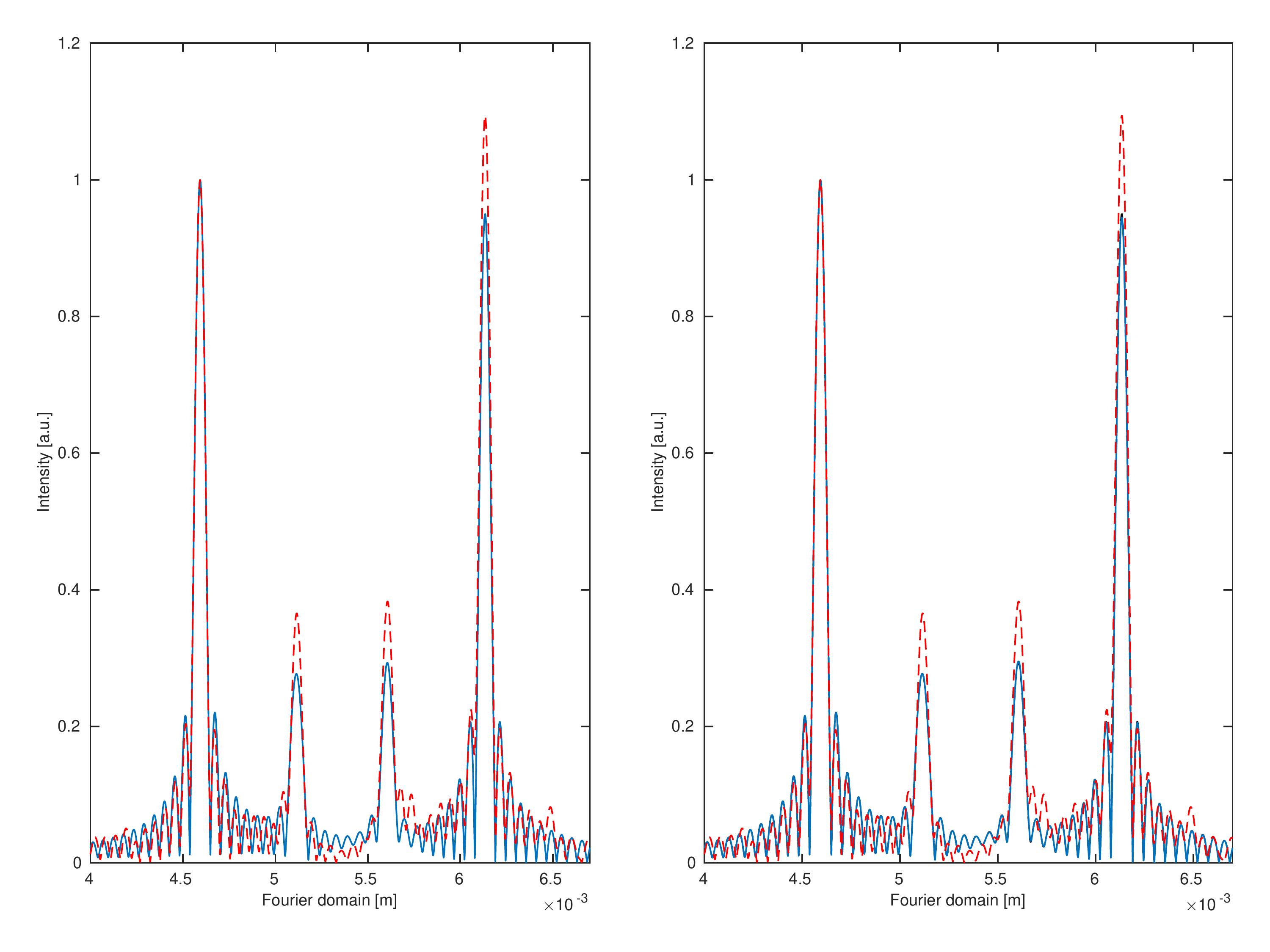}
\caption{Comparison between real data (red) and simulations using plane wave illumination (blue). We consider first- (left) and second-order (right) reflections. } 
\label{fig:real_vs_plane_multi}
\end{figure}

That this lack of information cannot be compensated by adding multiple reflections in the simulations is presented in Figure~\ref{fig:real_vs_plane_multi}, which shows a comparison, again in Fourier domain, between the real data and the plane wave model with first and multiple reflections. The multiple reflections were calculated up to second order. Hereby, we consider only the multiple reflections that travel at most the time needed for light to travel from the front to the back surface. 
The effect of the multiple reflections is minor as expected from the analytic representation. This is also true for higher order reflections. 

\subsubsection{Focusing Effect}

The effect showing responsible for this inconsistency between the two presented simulations in \autoref{fig:real_vs_plane_gaussian} is the focus of the laser beam. It is a strongly affecting factor that does not appear in the plane wave illumination but it is dominant in the Gaussian beam case. Thus, it is essential to consider this effect when we want to solve the inverse problem of reconstructing the material parameters from the OCT data. 

\begin{figure}[hbt!]
\centering 
\includegraphics[width=0.8\textwidth]{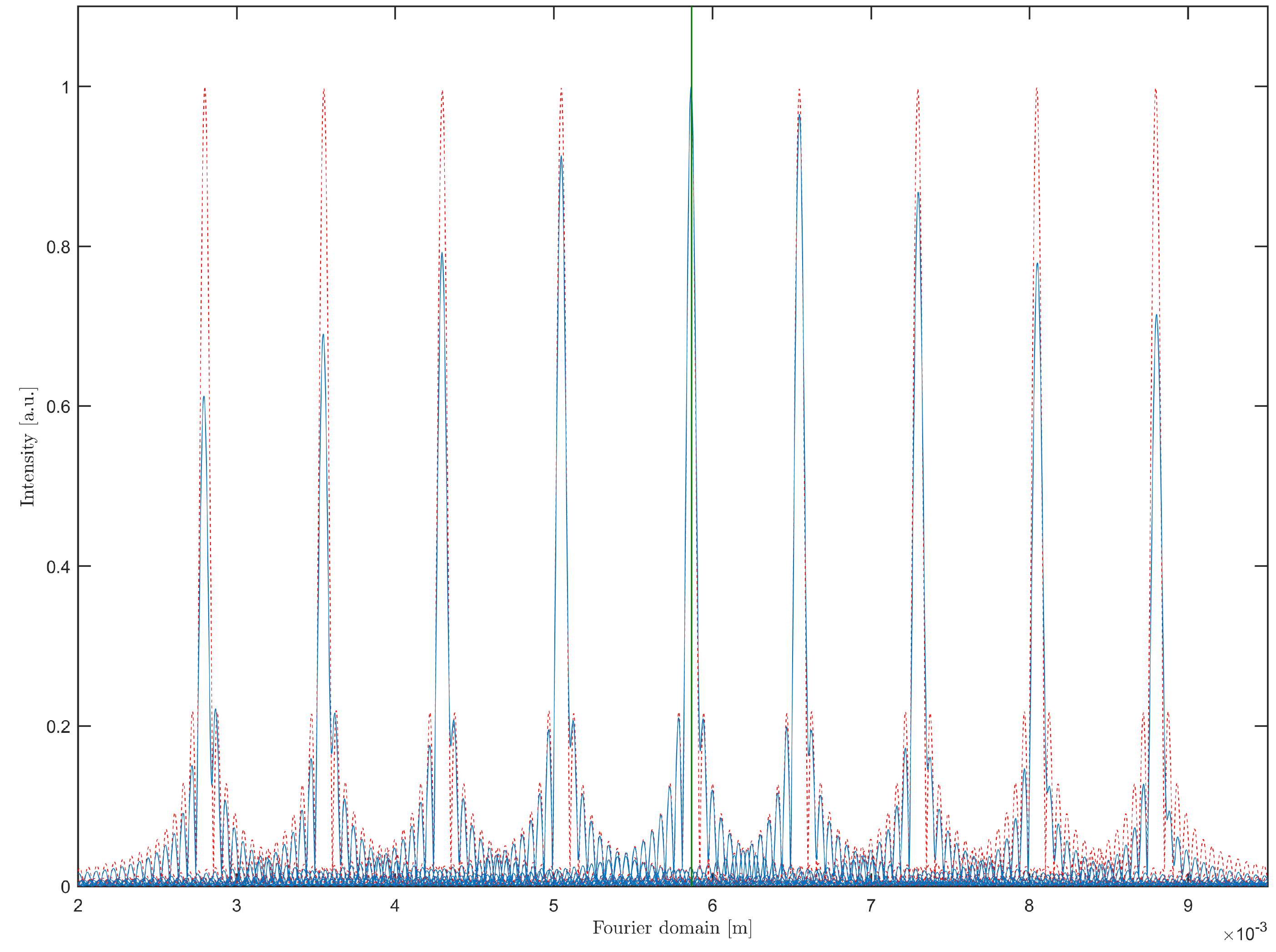}
\caption{Comparison of the plane wave model (red) and the Gaussian model for different object positions with respect to the focus (green line).} 
\label{fig:planefield}
\end{figure}

The difference between the plane wave and the Gaussian beam model is even more highlighted when the object of interest is moved out of the focus. Figure~\ref{fig:planefield} shows the absolute value of the Fourier transform of the cross-correlation term. Here, one can easily identify a strong decrease of the intensity as we move further from the focus position.
For the simulations, we assumed that the laser light is reflected by a non-tilted totally reflecting mirror. The mirror is shifted to different positions $x^j_\Omega,\, j=1,\dots,9,$ with respect to the focus position $r_3.$ For each position the cross-correlation term \eqref{eq:cc_term} is recorded on the exact same spectrum as in Section~\ref{sec:example_gaussian}.


\section*{Conclusions}
In this work we presented different possibilities for modeling an Optical Coherence Tomography system.  The performed analysis was based on different mathematical models originated from Maxwell's equations or their simplification, the Helmholtz equation. We tried to cover as many as possible cases considering various illumination techniques, medium properties, scattering behaviours and measurement ways. The addressed models were also tested/compared using simulated and real OCT data. 

\subsection*{Acknowledgements}
We would like to thank Wolfgang Drexler and Lisa Krainz both from the Medical University of Vienna for providing us with the experimental data presented in Section~\ref{sec:example_gaussian}. P.~Elbau and L.~Veselka acknowledge support from the Austrian Science Fund (FWF) within the project F6804–N36 of the Special Research Programme SFB F68 ``Tomography Across the Scales''.

\section*{References}
\renewcommand{\i}{\ii}
\printbibliography[heading=none]

\end{document}